\documentclass[sigconf,dvipsnames]{acmart}

\AtBeginDocument{%
  \providecommand\BibTeX{{%
    \normalfont B\kern-0.5em{\scshape i\kern-0.25em b}\kern-0.8em\TeX}}}
    
\usepackage[utf8]{inputenc}
\usepackage{url}
\usepackage{colortbl}
\usepackage{caption}
\usepackage{balance}
\usepackage{xspace}
\usepackage[caption=false]{subfig}
\usepackage{graphicx}
\usepackage{verbatim}
\usepackage{bold-extra}
\usepackage{amsmath}
\usepackage{multirow}
\usepackage{listings}
\usepackage{boxedminipage}
\usepackage{soul}
\usepackage{enumitem}
\usepackage[normalem]{ulem}
\setlist{nolistsep,leftmargin=.5cm}
\usepackage{booktabs}
\usepackage{tabularx}
\usepackage{svg}
\usepackage{diagbox}
\usepackage{calc}
\usepackage{float}
\usepackage{multirow}
\usepackage[normalem]{ulem}
\useunder{\uline}{\ul}{}
\usepackage{amsmath}
\usepackage[most]{tcolorbox}
\usepackage{caption}
\usepackage{microtype} 
\usepackage{cleveref}

\makeatletter

\newcommand{\BugLocator}{{\sc BugLocator}\xspace}
\newcommand{\SentenceBERT}{{\sc SentenceBERT}\xspace}
\newcommand{\UniXCoder}{{\sc UniXCoder}\xspace}
\newcommand{\Lucene}{{\sc Lucene}\xspace}
\newcommand{\BERT}{{\sc Bert}\xspace}

\newcommand{\TFIDF}{{\sc tf-idf}\xspace}

\definecolor{mycolor}{RGB}{204,204,204}

\newcommand*\circled[1]{\tikz[baseline=(char.base)]{\small{\textbf{
			\node[shape=circle,fill=mycolor,draw=black, inner sep=0.75pt] (char) {\textcolor{black}{#1}};}}}}

\newboolean{showcomments}

\setboolean{showcomments}{true}

\ifthenelse{\boolean{showcomments}}
  {\newcommand{\nb}[2]{
    \fbox{\bfseries\sffamily\scriptsize#1}
    {\sf\small$\blacktriangleright$\textit{#2}$\blacktriangleleft$}
   }
   
  }
  {\newcommand{\nb}[2]{}
   
  }

\newcommand*{\img}[1]{%
    \raisebox{-.1\baselineskip}{%
        \includegraphics[
        height=0.32cm,
        width=0.32cm,
        keepaspectratio,
        ]{#1}%
    }%
}

\newcommand{\ie}{\textit{i.e.,}\xspace}
\newcommand{\eg}{\textit{e.g.,}\xspace}
\newcommand{\etc}{\textit{etc.}\xspace}
\newcommand{\etal}{\textit{et al.}\xspace}

\copyrightyear{2024}
\acmYear{2024}
\setcopyright{rightsretained}
\acmConference[ICSE '24]{2024 IEEE/ACM 46th International Conference on Software Engineering}{April 14--20, 2024}{Lisbon, Portugal} 
\acmBooktitle{2024 IEEE/ACM 46th International Conference on Software Engineering (ICSE '24), April 14--20, 2024, Lisbon, Portugal}
\acmDOI{10.1145/3597503.3608139} 
\acmISBN{979-8-4007-0217-4/24/04}

\usepackage{xspace}

\begin{document} 

\title{
	On Using GUI Interaction Data to Improve \\ Text Retrieval-based Bug Localization
}

\author{Junayed Mahmud}
\email{junayed.mahmud@ucf.edu}
\affiliation{%
	\institution{University of Central Florida}
	\country{Orlando, FL, USA}
}

\author{Nadeeshan De Silva}
\email{kgdesilva@wm.edu}
\affiliation{%
	\institution{William \& Mary}
	\country{Williamsburg, VA, USA}
}

\author{Safwat Ali Khan}
\email{skhan89@gmu.edu}
\affiliation{%
	\institution{George Mason University}
	\country{Fairfax, VA, USA}
}

\author{Seyed Hooman Mostafavi}
\email{smostaf6@gmu.edu}
\affiliation{%
	\institution{George Mason University}
	\country{Fairfax, VA, USA}
}

\author{SM Hasan Mansur}
\email{smansur4@gmu.edu}
\affiliation{%
	\institution{George Mason University}
	\country{Fairfax, VA, USA}
}

\author{Oscar Chaparro}
\email{oscarch@wm.edu}
\affiliation{%
	\institution{William \& Mary}
	\country{Williamsburg, VA, USA}
}

\author{Andrian Marcus}
\email{amarcus7@gmu.edu}
\affiliation{%
	\institution{George Mason University}
	\country{Fairfax, VA, USA}
}

\author{Kevin Moran}
\email{kpmoran@ucf.edu}
\affiliation{%
	\institution{University of Central Florida}
	\country{Orlando, FL, USA}
}

\renewcommand{\shortauthors}{J. Mahmud, N De Silva, S.A. Khan, S.H. Mostafavi, SM. H. Mansur, O Chaparro, A. Marcus, and K. Moran.}

\begin{abstract}

One of the most important tasks related to managing bug reports is \textit{localizing the fault} so that a fix can be applied. As such, prior work has aimed to automate this task of bug localization by formulating it as an information retrieval problem, where potentially buggy files are retrieved and ranked according to their textual similarity with a given bug report. However, there is often a notable \textit{semantic gap} between the information contained in bug reports and identifiers or natural language contained within source code files. For user-facing software, there is currently a key source of information that could aid in bug localization, but has not been thoroughly investigated -- information from the graphical user interface (GUI).

In this paper, we investigate the hypothesis that, for end user-facing applications, connecting information in a bug report with information from the GUI, and using this to aid in retrieving potentially buggy files, can improve upon existing techniques for text retrieval-based bug localization. To examine this phenomenon, we conduct a comprehensive empirical study that augments four baseline text-retrieval techniques for bug localization with GUI interaction information from a reproduction scenario to (i) filter out potentially irrelevant files, (ii) boost potentially relevant files, and (iii) reformulate text-retrieval queries. To carry out our study, we source the current largest dataset of fully-localized and reproducible real bugs for Android apps, with corresponding bug reports, consisting of 80 bug reports from 39 popular open-source apps. Our results illustrate that augmenting traditional techniques with GUI information leads to a marked increase in effectiveness across multiple metrics, including a relative increase in Hits@10 of 13-18\%. Additionally, through further analysis, we find that our studied augmentations largely complement existing techniques, pushing additional buggy files into the top-10 results while generally preserving top ranked files from the baseline techniques.

\end{abstract}
\vspace{-2em}

\vspace{-3em}

\maketitle

%-------------------------------------
\vspace{-1em}
\section{Introduction}

The process of bug report management has been demonstrated to consume large amounts of developer's time \cite{murphy2013design,weiss2007long}. One of the more difficult bug management tasks is related to \textit{localizing the described fault}, as it requires reasoning between the description of a bug and the source code of a software project. 
This process is often further complicated by quality issues related to various elements of bug descriptions, such as reproduction steps, stemming from inaccurate or incomplete information provided by reporters~\cite{Chaparro2019,Bettenburg2008GoodBR,Chaparro2017,song2020bee}.

Researchers have been working to automate \textit{bug localization} by developing approaches that automatically retrieve and rank potentially buggy files or code snippets to help expedite localization effort. 
A substantial body of research formulates bug localization as a text retrieval-based (TR) problem \cite{davies2012using} --- see \Cref{sec:related_work}. In general, these approaches use the bug report to formulate a query and return a list of source code elements (files, classes, methods, \etc), ranked by their likelihood that they contain the bug.

The key assumption made by TR-based bug localization approaches is also their biggest limiting factor. 
That is, while such techniques operate on the premise that bug reports and the corresponding buggy source code will share terms, research has also documented a notable \textit{semantic gap} between the information that reporters provide in bug reports, and the identifiers and the documentation written by developers in source code~ \cite{Zimmermann2010,marcus2001identification,Moran2015}. 
Researchers have recognized this issue, and have attempted to augment TR-based bug localization approaches with various techniques.
Many approaches focused on processing the text in the bug reports or the source code (\eg through abbreviation expansion \cite{hill2008amap}), while others focused on query reformulation, or automatically augmenting a query generated from a bug report using information various sources \cite{Rahman2018}.
Another line of research focused on using information orthogonal to the code and bug report vocabulary to boost the ranks of the retrieved buggy code elements, such as, execution information (extracted form execution or stack traces \cite{youm2017improved, Wen2016}), code dependencies (extracted via static source code analysis \cite{dit2013feature}), or historical information (extracted from repositories \cite{youm2017improved}).

\begin{figure*}[t]
		\vspace{-0.2cm}
		\centering
		\includegraphics[width=0.95\linewidth]{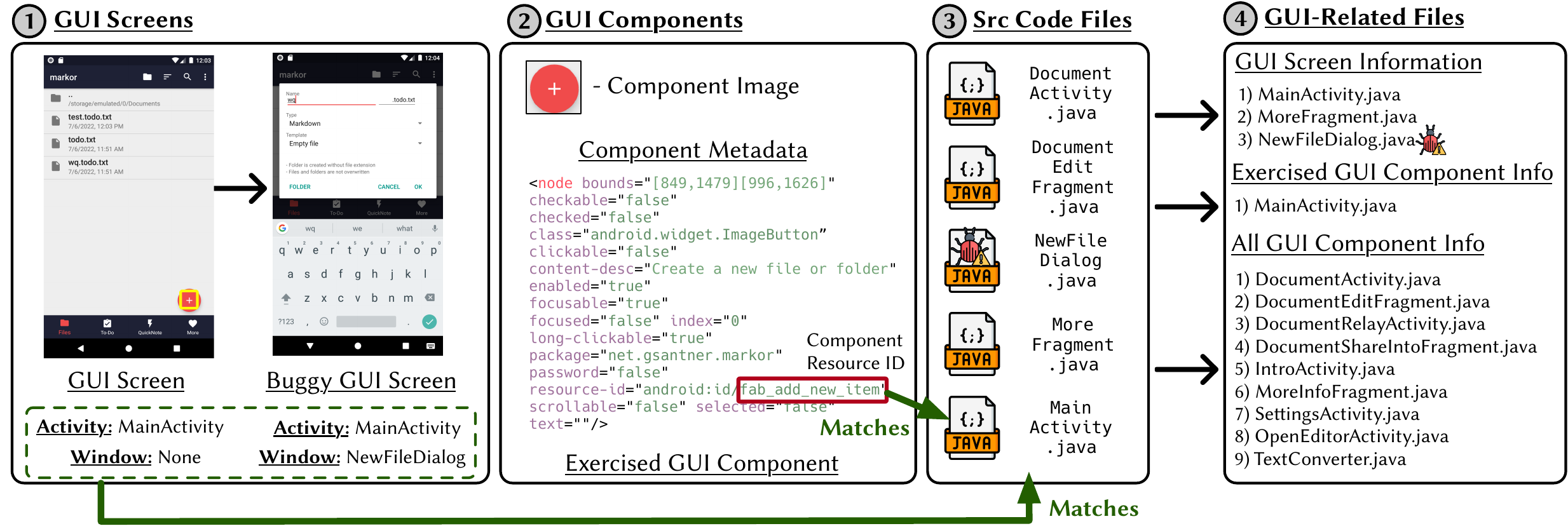}
		\caption{Example of GUI-related information used in this study}
		\label{fig:def-examples}
		\vspace{-1em}
\end{figure*}

In this paper, we explore whether it is possible to improve TR-based bug localization leveraging an information source not yet explored by prior work  -- \textit{information from the graphical user interface (GUI)}. {GUIs encode latent patterns related to application features in both pixel-based (\ie screenshots) and  metadata-based (\ie \texttt{\small html/uiautomator}) representations~\cite{Moran:SANER22}.} Our rationale 
is that GUI interaction information can be easily obtained and represents "high-level" execution related information, where code elements are directly linked to higher level program functionality through the UI, as opposed to "low-level" execution information extracted from execution traces, which can be difficult to acquire. 
Once collected, this high-level GUI information can then be used to boost rankings of related buggy code elements.
Further, unlike low-level execution traces, GUI-related information is rich in textual elements, and can also be used for query reformulation.
Intuitively, if the buggy code is \textit{related} to the app screen where the bug is observed, then the GUI and interaction information from that screen can be used to help locate the buggy code easier. Conversely, if the buggy code is not related to the buggy screen or user interactions, then we expect that the GUI-related information will not hinder the bug localization process. 
We refer to the collective data related to both user interactions and the software interface itself as \textit{GUI interaction~data}.
\looseness=-1

To investigate whether GUI interaction data can aid in TR-based bug localization, we carry out a comprehensive empirical study that \textit{augments} four baseline TR-based approaches: BugLocator~ \cite{Zhou2012}, a Lucene-based approach ~\cite{apacheLucene}, and two neural-based text embedding approaches (based upon the sentenceBERT \cite{reimers2019sentence} and UniXCoder~\cite{guo2022unixcoder} neural language models). 
The GUI information that we use for augmentation is collected from a recorded set of GUI interactions that reproduce a given bug, which can be easily collected manually by developers, or automatically by any of a number of bug reproduction techniques~\cite{Johnson2022,Fazzini2018,Zhao2019}. 
Once these GUI interactions are collected, we assess the effect on retrieval performance by using information from the GUI to: 
(i) filter out potentially irrelevant files that are not related to the buggy GUI screen; 
(ii) boost potentially relevant files that are related to the buggy GUI screen; 
and (iii) reformulate queries using information from the buggy GUI screen. 
\looseness=-1

In this study, we focus on localizing bugs in Android apps, which typically manifest themselves in the GUI. 
This means that these bugs lead to unexpected app behavior (or a faulty state) that is visible to the user, including app crashes (\eg when the app suddenly closes), navigation issues (\eg when the app leads the user to an unexpected screen), and incorrect output  shown on the screen, among others. 
As such, to support our study, we have manually sourced and validated the current largest dataset of fully-localized and reproducible bugs for Android applications with corresponding bug reports, consisting of 80 bug reports from 39 popular open source Android applications. We compared our baseline TR-based bug localization approaches with thousands of augmented configurations using different types and amounts of GUI interaction data, as well as different query reformulation techniques. 

Our results illustrate the benefit of leveraging GUI interaction information, as the best-performing configurations of the techniques augmented with GUI information outperformed their baseline for \textit{all} TR-based techniques, with Hits@10 improving by 13-18\% overall. 
A deeper investigation into these results show that our studied augmentations help rank more bugs in the top-10 retrieved results, while generally preserving the top-ranked buggy files from the baseline techniques. 
Overall, the results support our rationale for leveraging GUI information to improve bug localization, and suggest that future work should explore this topic further.
\looseness=-1

In summary, this paper makes the following contributions:

\begin{itemize}
	\item{A new dataset of 80 fully-localized and reproducible Android bugs from 39 popular Android apps, complete with bug reports and recorded scenarios (and metadata) that reproduce each bug,}
	\item{A quantitative analysis of the effect of using GUI interaction information on the effectiveness of four TR-based bug localization techniques, via three augmentation methods: (i) filtering, (ii) boosting, and (iii) query reformulation}, and 
	\item{A replication package~\cite{faultLocalisationCode} that contains our dataset, code, and experimental infrastructure to aid in the replication and reproduction of our results.}
\end{itemize}

\vspace{-1em}
\section{Background \& Motivation}
\label{sec:motivation}

In this section, we provide background on the various GUI-related terms and concepts related to our study, explain our methods for augmenting text-retrieval techniques for bug localization with GUI-related information, and include a motivating scenario that illustrates the intuition behind the augmentation methods.

\subsection{GUI-related Information}

In this paper, we analyze GUI information of Android apps, namely \textbf{GUI screens},  \textbf{GUI components} (\ie the GUI widgets/elements that compose those screens), and \textbf{Exercised Components} (\ie the GUI components that the user interacts with via taps, swipes, \etc). 
To illustrate the definitions of these various GUI-related concepts, we provide an example in Figure~\ref{fig:def-examples}, oriented around a set of screens that reproduce the bug described in the report shown in Figure~\ref{fig:bug-example}.

\noindent\circled{1}~\textbf{\uline{GUI screens}} represent the UI canvas upon which \textbf{GUI components} are drawn, wherein the screen is composed of a hierarchy of interactive components and GUI containers that group individual components together such that they may adapt to various screen sizes and dimensions. 
In Android, \textbf{screens} are referred to as Activities~\cite{activities}, and each activity corresponds to one or more \texttt{\small .java}/\texttt{\small .kotlin} class files that define the functionality of the screen, and a set of resource \texttt{\small .xml} files that describe the layout of components on the screen.
 The code and resource files directly make up the static definition of the screen in the code.
In addition to Activities, Android also allows for the definition of Fragments~\cite{fragments}, which are reusable groups of \textbf{GUI components} (\eg menus, dialog boxes...). 
A \textbf{GUI screen} can display a Window, as shown in the \textit{buggy screen} of Figure~\ref{fig:def-examples}-\circled{1}, which can display a dialog, toast, or other \textbf{GUI component} in the foreground of the Activity.  
Fragments and Windows are also defined in their own \texttt{\small .java}/\texttt{\small .kotlin} class files and \texttt{\small .xml} resource files.
\looseness=-1

In addition to the static definitions of \textbf{GUI screens} in class and resource files, it is also possible to extract a runtime representation of a given \textbf{GUI screen} using the Android \texttt{\small uiautomator} framework~\cite{androidUiAutomator}, which queries a device's \texttt{\small ViewServer} to extract metadata about the various \textbf{GUI components} currently rendered on the screen.
Figure~\ref{fig:def-examples} illustrates the last two \textbf{screens} of the bug reproduction for the Markor app \cite{markorApp}, along with their Activity (\texttt{\small MainActivity}), and Window (\texttt{\small NewFileDialog}) information. 
Note that both screens correspond to the same Activity, but the second screen displays a foreground popup defined in the \texttt{\small NewFileDialog} Java class file.

\noindent\circled{2}~\textbf{\uline{GUI Components}} are the UI elements, defined by developers and rendered to the screen, with which end-users interact via GUI-level actions (taps, swipes, long touches, \etc). 
The presentation attributes (color, size, component type, \etc) are defined in the app source code and in the resource files described earlier. These component definitions can be attached to \texttt{\small Event Listeners} that cause Java/Kotlin code to be executed when a certain GUI-level action (\eg tap) is performed on the component. 
To link the \textbf{component} definition to the app Java/Kotlin code, unique Resource IDs are used. 
As mentioned earlier, in addition to the static definition of GUI components, a dynamic representation can also be extracted using the \texttt{\small uiautomator} framework. 
This dynamic \textbf{component} metadata is shown as in Figure~\ref{fig:def-examples}-\circled{2} for the \img{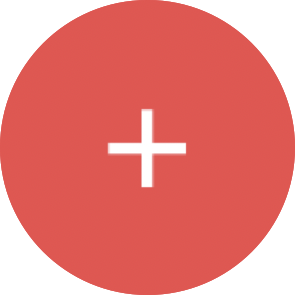} button in the bottom right hand side of the first GUI Screen (highlighted in yellow). 
This metadata contains information related to the \texttt{\small size}, \texttt{\small class}, and \texttt{\small resource-id} of this \textbf{component}, among other attributes. 
If you refer to the screen flow shown in Figure~\ref{fig:def-examples}-\circled{1}, the \img{images/plus.png} button was the \textbf{component} that the user interacted with to navigate to the buggy \textbf{screen} with the dialog box. 
We refer to \textbf{components} that a user interacted with during a GUI-level execution scenario as \textbf{Exercised Components}.

\vspace{-1em}
\subsection{Motivating Example}

\begin{figure}[t]
	\vspace{-0.1cm}
	\centering
	\includegraphics[width=\linewidth]{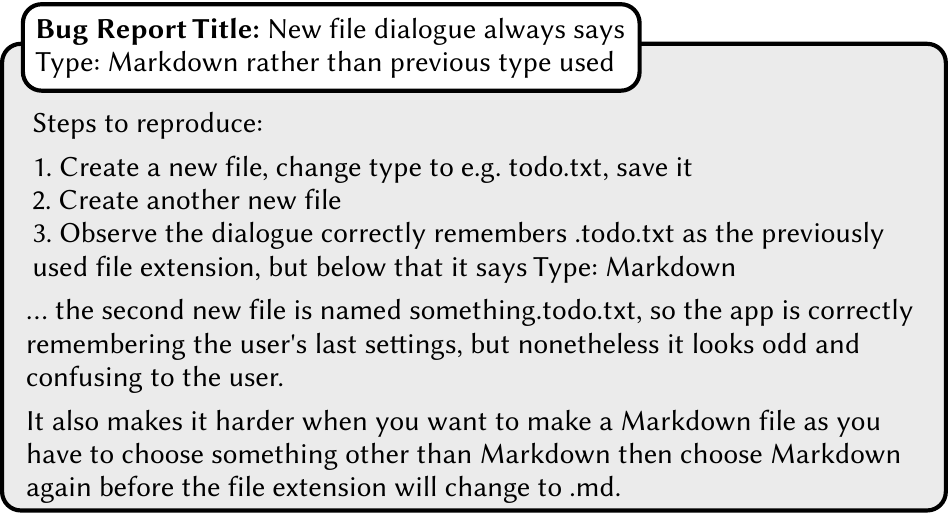}
	\caption{Example Bug Report from the Markor\cite{markorBugFig1} App}
	\label{fig:bug-example}
	\vspace{-0.6cm}
\end{figure}

In order to illustrate our intuition about leveraging GUI interaction information to improve bug localization, 
in this subsection we illustrate the effect that this information has on an example bug report 
from the Markor~\cite{markorBugFig1} app shown in Figure~\ref{fig:bug-example}. 

The Markor app is a text-editor primarily aimed at supporting quick note-taking and managing to-do lists. It is also relatively popular -- the GitHub repository currently has 2.6k stars. 
For each file created, Markor supports different ways of formatting the text (\eg Markdown) that is configurable through a note creation dialog. 
The observed buggy behavior of the report (illustrated in Figure~\ref{fig:bug-example}) is related to the formatting method of a given file being reset in the File creation/editing dialog, wherein it always defaults to ``Markdown'', even when it is initially set to something else. 
For instance, if a user were to set the note to ``plain-text'' formatting, and then later return to edit the note, the formatting type would be reset to ``Markdown''.
\looseness=-1

Let's consider an existing TR-based bug localization approach, BugLocator~\cite{Zhou2012}, which uses term similarity and document length in conjunction with a Vector Space Model (VSM), to rank buggy files. 
When we use the bug report from Figure~\ref{fig:bug-example} to BugLocator as a query, the top 3 returned files are (1) \texttt{\small SearchOrCustomText}- \texttt{\small DialogCreator.java}, (2) \texttt{\small TodoTxtHighlighter.java}, and (3) \texttt{\small TextFor-} \texttt{\small mat.java}. However, as shown in Figure~\ref{fig:def-examples}-\circled{3} the actual buggy file is \texttt{\small NewFileDialog.java}, as the bug occurs in the Dialog box used for creating and modifying files. 
This file is ranked \textit{\textbf{35th}} by BugLocator. 
However, for this particular bug, we can observe in Figure~\ref{fig:def-examples}-\circled{1} that the screen which exhibits the buggy behavior is composed of the \texttt{\small MainActivity} activity, and the \texttt{\small NewFileDialog} fragment that makes up the popup dialog box shown in the screenshot. 
For this particular bug, \textit{the java class file \texttt{\small NewFileDialog.java} that corresponds to the window fragment in the buggy screen \uline{is the buggy file}} (shown in  Figure~\ref{fig:def-examples}-\circled{3}). 
Hence, if the TR-based bug localization approach had knowledge of this GUI-related information, it likely could have improved the ranking of the buggy \texttt{\small NewFileDialog.java} file.

This example illustrates the potential promise of leveraging GUI interaction data to improve existing text retrieval-based bug localization approaches.
While the buggy file for a given report may not always correspond neatly to the class that implements a buggy screen's Activity or Window, information related to event-listeners for various GUI-components that are relevant to a bug reproduction scenario, or screen information from earlier bug reproduction steps, may help identify and rank buggy files. 
In this study, we aim to investigate the potential for such \textit{GUI Interaction Information} to augment   TR-based bug localization approaches.

\vspace{-1em}
\section{Design of Empirical Study}

The goal of this study is to investigate: (1)  the extent to which GUI interaction information from Android apps leads to more effective automated bug localization, and (2)  how this information can be most effectively used to increase bug localization effectiveness. With this in mind, we formulate the following overarching research question (RQ), refined into four specific research RQs:

\begin{enumerate}[label=\textbf{RQ:}, ref=\textbf{RQ}, wide, labelindent=0pt]\setlength{\itemsep}{0.3em}
	\item\label{rq:gui-information}{\textbf{\textit{What is the impact of using GUI interaction information on text-retrieval bug localization performance?}}}
		\begin{itemize}\setlength{\itemsep}{0.3em}
			\item{\textbf{RQ$_{1}$:} \textit{What is the effect of using GUI interaction information from different numbers of screens on bug localization performance?}}
			\item{\textbf{RQ$_{2}$:} \textit{What is the effect of the type of GUI interaction information and augmentation method on bug localization performance?}}
			\item{\textbf{RQ$_{3}$:} \textit{What are the overall best-performing combinations of using GUI interaction information?}}
		\end{itemize}
\end{enumerate}

To answer these research questions, we have developed methods to link GUI Information to potentially buggy files and to augment existing TR-based bug localization techniques with this file information, which we describe in Secs.~\ref{subsec:gui-mapping} and~\ref{subsec:augmentations}. 
Additionally, we have manually sourced the largest corpus of fully-localized and reproducible Android bugs, consisting of 80 bug reports from 39 popular open source Android apps (see Sec.~\ref{sec:dataset}).

We study four \textit{baseline} techniques -- one existing TR-based bug localization technique, one traditional term matching technique, and two neural text-embedding approaches, described in Sec.~\ref{sec:baselineTechniques}. We assess the performance of these baseline techniques in locating buggy files for the 80 bugs, with and without using GUI-information augmentation. We measure bug localization performance using the metrics described in Sec.~\ref{sec:metrics}, and answer the RQs in Sec.~\ref{sec:results}.

\subsection{Mapping GUI Terms to GUI-Related Files}
\label{subsec:gui-mapping}

Given that the goal of our study is to determine how GUI information can be used to augment bug report-based fault localization, we aim to map \textbf{terms} extracted from \textbf{GUI Screens} and \textbf{Components} to source code \textbf{files} that may be useful for the localization process. These \textbf{GUI-related terms} are later used for query reformulation, whereas the \textbf{GUI-Related Files} are used to re-rank retrieved files.
Our study assumes the scenario wherein the developer has access to a GUI-level reproduction scenario, which contains screenshots, the uiautomator metadata, and the \texttt{\small resource-id} and GUI-level action (\eg tap, long touch) for each screen interaction. 
This information is easy to collect, and could be collected manually, wherein the developer records a video and translates this video to a scenario using a tool such as V2S \cite{Havranek2021,bernal2023translating,bernal2020translating} or GifDroid \cite{Feng2022}, or can be collected 
automatically using tools that reproduce Android bug reports such as Yakusu \cite{Fazzini2018}, RecDroid+ \cite{Zhao2019,zhao2022recdroid+}, or the recent approach by Zhang \etal~\cite{zhang2023automatically}. 
Given this information, we aim to link \textit{key terms} from \textbf{GUI-screens} and \textbf{GUI-components} to potentially buggy files, or \textbf{GUI-Related Files}. 
Given that the bug occurs at the end of a given reproduction scenario, but the buggy behavior may be triggered or exercised earlier in the scenario, we explore using information from the buggy \textbf{screen} and the information from the prior 1-3 \textbf{screens}. 
Next, we describe how we identify \textbf{GUI-Related Files} from \textbf{GUI Screen} and \textbf{component} metadata (using the relevant \textbf{terms}).
\looseness=-1

\newcolumntype{P}[1]{>{\centering\arraybackslash}p{#1}}
\begin{table}[t]
\footnotesize
\caption{Mapping of GUI Terms to GUI-Related Files}
\label{tab:gui-files}
\begin{tabular}{P{2.5cm}|P{2.5cm}|P{2.5cm}}
\multicolumn{1}{c|}{\textbf{GUI Information}} & \textbf{Terms}                                                             & \textbf{Files}                                                                          \\ \hline
Screen                                        & Activity and Window names queried from the Android \texttt{ViewServer}                     & \texttt{\footnotesize .java} files with file names matching the terms                              \\ \hline
(Exercised) GUI Screen Components             & \texttt{resource-id(s)} of the (interacted) components from the dynamic \texttt{uiautomator} metadata & \texttt{\footnotesize .java}  files that contain invocations of the \texttt{resource-id(s)} in event listeners \\ \hline
\end{tabular}
\vspace{-2em}
\end{table}

We define three different types of \textbf{GUI-Related Files}: 
(1) those related to the Activity and Window information from a given \textbf{GUI Screen}; 
(2) those related to the \textbf{Exercised Components}; 
and (3) those related to all the components that appear on the selected screens of a given reproduction scenario, which we refer to as \textbf{GUI Screen Components}. 
We next discuss how each type of \textbf{GUI-Related File} is derived, also summarized in Table~\ref{tab:gui-files}.

\noindent\textbf{\uline{- Mapping GUI Screen terms to Files:}} To derive the GUI-Related Files for a given \textit{GUI screen}, the Activity and Window names from the dynamic UI metadata generated by the Android \texttt{\small ViewServer} are taken as terms and directly matched with their corresponding \texttt{\small .java} class file names. 
For instance, the \textit{GUI Screen} information for the two screens shown in Figure~\ref{fig:def-examples} is \texttt{\small MainActivity}, \texttt{\small MoreFragment}, and \texttt{\small NewFileDialog} as shown in Figure~\ref{fig:def-examples}-\circled{4}, which map to the Java files with the same names.

\noindent\textbf{\uline{- Mapping GUI Screen Component terms to Files:}} To derive GUI-Related Files for \textit{all} GUI components of the considered screens,  the \texttt{\small resource-id} for all components for a given screen that support interaction are extracted from the \texttt{\small uiautomator} metadata and used as terms, and then invocations of these \texttt{\small resource-id} terms are automatically identified in event-listeners in the app source code. The corresponding files that contain these event listeners are taken as the GUI-Related Files for Screen Components. This generally leads to a large set of GUI-Related Files, as matching the \texttt{\small resource-ids} of a large number of interactive components on a given set of screens to event-listeners typically leads to many affected files. 
For the example shown in Figure~\ref{fig:def-examples}, the \texttt{\small resource-ids} for 41 components are extracted, which in turn map to event-listeners in 77 code files, some of which are shown in Figure~\ref{fig:def-examples}-\circled{4}.

\noindent\textbf{\uline{- Mapping Exercised GUI Component terms to Files:}} 
To derive GUI-Related Files for an \textit{Exercised GUI Component}, the same approach as described for GUI Screen Components is followed, but \textit{only} for those components on the screen which were exercised as part of a bug reproduction scenario.
Any source code file that contains an event-listener for the component is identified is considered as a GUI-Related File. 
For the example in Figure~\ref{fig:def-examples}, there is only one Exercised Component, and only one invocation of this exercised component, in the \texttt{\small MainActivity.java} file (shown in Figure~\ref{fig:def-examples}-\circled{4}).

It is important to note that in our study we do not only consider each type of GUI information in isolation, but also consider combinations, represented as unions of the GUI Information Terms for two different types. 
We illustrate our considered combinations in Table~\ref{tab:gui-info}, organized according to the number of files they typically return. 
Note that GUI Information terms stemming from Exercised GUI Components are a subset of the GUI Screen Components, thus we do not combine these information types.
Additionally we ignore components without \texttt{\small resource-ids} and do not consider interactions with the Android \texttt{\small Back} button from the bottom navigation bar as exercised components, as interactions with this component cannot be mapped back to the event listeners in the app code.
\looseness=-1

\begin{table}[t]
\footnotesize
\vspace{-0.5em}
\caption{Types of GUI-Related Files}
\begin{tabular}{l}
\hline
\textbf{Low Number of Files}                  \\ \hline
\enspace GUI Screen                                    \\
\enspace Exercised GUI Components                      \\ \hline
\textbf{Medium Number of Files} 				\\ \hline
\enspace GUI-Screen + Exercised GUI Components         \\ \hline
\textbf{High Number of Files}                 \\ \hline
\enspace GUI Screen Components                            \\
\enspace GUI Screen + GUI Screen Components               \\ \hline
\end{tabular}
\label{tab:gui-info}
\vspace{-2em}
\end{table}

\subsection{Text-Retrieval Augmentation Methods}
\label{subsec:augmentations}

Below, we describe how the GUI-Related Files can be used to augment existing techniques for text retrieval via Query Reformulation and Re-Ranking, to (potentially) improve bug-localization. 

In the context of our bug localization process, typically text-retrieval techniques use a pre-processed version of the bug report as a query to retrieve source code files that contain similar terms to those used in the bug report. 
The method of calculating query-to-document similarity can vary from relatively simple methods, such as using term frequency (\eg \texttt{\small tf-idf} vectors), to more complicated techniques that use neural text embeddings.

\subsubsection{\textbf{Reformulating Queries using GUI Terms}}  
The first type of augmentation techniques that we define are \textbf{Query Reformulation} techniques~\cite{florez2021combining}.
These techniques modify the textual query used by TR-based techniques to retrieve relevant files. Given this setting, we define the following two reformulation techniques: 

\noindent\textbf{\uline{- Query Expansion:}} In this technique, the dynamic Activity/Window names for GUI screens, and GUI component resource IDs for a given GUI component type (\eg Exercised Components or Screen Components) are appended to the bug report to form the query.  
\looseness=-1

\noindent\textbf{\uline{- Query Replacement:}} In this technique, the dynamic Activity/Window names for GUI screens and GUI component resource ids for a given GUI component type (\eg Exercised Components or Screen Components) are used to replace the bug report as the query.
\looseness=-1

\subsubsection{\textbf{Re-Ranking using GUI-Related Files}} 

We explore three different techniques for \textbf{re-ranking} files using GUI information:

\noindent\textbf{\uline{- Filtering:}} In this strategy, all files that \textit{do not} match the GUI-Related Files for a given information type are filtered out from the corpus of potentially buggy files. 

\noindent\textbf{\uline{- Boosting:}} In this strategy, files that \textit{match} the GUI-Related Files are boosted to the \textit{top} of the ranked list of results returned by a given text-retrieval technique while preserving the relative order of those files originally ranked by the technique.

\noindent\textbf{\uline{- Filtering + Boosting:}} The final re-ranking strategy combines both filtering and boosting, wherein files are filtered using a GUI information type that has a higher number of files, and boosting is performed with a type that returns a lower number of files. 
This is due to the fact that filtered files, cannot be subsequently boosted.

\vspace{-1em}
\subsection{Dataset Construction}
\label{sec:dataset}

\subsubsection{\textbf{Bug Report Selection}}
\label{sec:bugLocalizationDataset}

Given that no prior dataset of fully-localized and reproducible bug reports for Android apps exists, we constructed our own using a rigorous manual process. 
We built a \textit{ground-truth} bug localization dataset by using as many bug reports as possible from the AndroR2 dataset~\cite{Wendland2021,Johnson2022}, which consists of 180 manually reproduced bug reports for popular open source Android applications hosted on GitHub. 
These reports were systematically collected from the project's issue trackers according to the following criteria, as reported by Wendland \& Johnson~\cite{Wendland2021,Johnson2022}:  they contain the label "bug", were opened in the past five years and closed at the time of the mining (November 2020), contain the word "steps" in their content, and report a non-trivial bug (\ie did not occur by opening the app). 
The bug reports are grouped into four categories that represent a bug type, namely \textit{output-}, \textit{cosmetic-}, \textit{navigation-}, and \textit{crash-}related bugs. 
Furthermore, each bug report is associated with additional (meta)data, including the commit ID of the app version that contains the bug, the buggy app's \texttt{\small apk} file, and the link to the GitHub issue where it was originally submitted. 

We explored all bug reports in the AndroR2 dataset and utilized a subset of reports that fit the necessary criteria for our study (\ie they were reproducible and able to be localized to source code files). To identify the buggy files among the set of all files from the buggy version of the app, we followed a systematic procedure that involved at least two authors examining each bug report. Specifically, two authors inspected the content of the bug reports (including the comments) to find any references to the commits that fixed the bug. 
They then inspected the commit messages and specific code changes to determine if they appeared to fix the bugs. Among the 180 bug reports in AndroR2, one bug report does not exist anymore, and eight bug reports do not contain any obvious commit ID or version information for the bug being reported, making it difficult to extract the buggy source code.
As a first step in filtering the dataset, we excluded these bug reports and investigated the remaining 171 bug reports.
\looseness=-1
 
In checking for bug-fixing commits, the two authors were able to source 120 bug reports with bug-fixing commit IDs that were confirmed to fix the reported bug, and on average, each bug report contained $\approx$1.90 bug-fixing commit IDs.
When no commit was mentioned in the bug reports, or the fixed commit ID mentioned in bug reports did not appear to resolve the error, the two authors followed references to duplicate reports and collected the referenced commits, again confirming that these commits did indeed fix the bug. 
We followed this procedure for the remaining 51 bug reports (of the 171 filtered reports) and were successful in gathering bug-fixing commit IDs in 27 cases -- resulting in 147 bug reports with confirmed bug-fixing commit IDs. 
Although in certain cases our ground-truth uses the bug fixing commits from duplicate bug reports, in this study we use the original bug report contents as reported by AndroR2 as the query for TR-based bug localization.
Once bug-fixing commits were identified, two authors performed two rounds of coding, during which they compared the file diffs between the buggy (i.e., the latest app release commit ID before the bug was reported) and the fixed version of the app (based on the commit IDs) to identify the files that contained bug fixing changes. If there were disagreements, a third author discussed the cases with the two coders and the three authors reached a consensus.
Note that this process only identified code files from the app's buggy version that had bug fixes (\ie code changes), thus excluding code files that were \textit{added} in the fixed app version and ignoring changes in white space and code comments. This set of 147 bug reports formed the set used for isolating the buggy files and collecting the required bug reproduction scenarios and GUI information.

\subsubsection{\textbf{Coding \& Collecting GUI Interaction Data}}
\label{sec:coding}
The ground-truth construction process was executed independently by two authors across two sessions. 
During the first session, two authors randomly selected 24 bug reports from all failure types of the AndroR2 dataset among the 147 bug reports filtered as described above. 
The two authors worked together in this session and discussed each bug in order to derive a common understanding of the coding process. 
Given that some of the studied baseline techniques \textit{only} operate on \texttt{\small .java} files, we discarded two bugs where the erroneous behavior was isolated the \texttt{\small .xml} resource files and one bug containing buggy Kotlin files. 
As such, 21 bug reports were coded as part of this first session. 

To collect the GUI interaction data for these 21 bug reports, we followed the record-and-replay methodology used by Cooper~\etal \cite{Cooper2021}, which includes installing and using the buggy app on an emulator, recording a usage scenario that reproduces the bug while collecting \textit{screen recordings} and \texttt{\small getevent} traces (\ie traces that include low-level GUI-related information) using the Android \texttt{\small getevent} utility. 
These scenarios are then replayed in a step-by-step manner by converting the recorded \texttt{\small getevent} actions to a set of \texttt{\small adb} commands. 
During the step-by-step replay, screenshots and GUI metadata were collected before and after each GUI action that includes information on the Activity and Window, as well the GUI hierarchy extracted via the \texttt{\small uiautomator} tool that contains the UI metadata for all components displayed on the screen. Since the AndroR2 bug reports provide the steps to reproduce the bug (S2Rs) and the buggy APK of the apps, we were able to reproduce the bugs on an emulator. 
Specifically, two authors reproduced the bugs manually on Pixel 2 Android emulator with the specific Android version mentioned in the AndroR2 dataset. 
Among the 21 bug reports, we could not reproduce the bug for one bug report due to a lack of \texttt{\small getevent} support for recording rotation events. Therefore, we  included the remaining 20 bug reports in our dataset.

In the second round of coding, the two authors worked with the remaining 123 bug reports. During this process 16 additional bug reports were identified to have bug fixing changes  on \texttt{\small .xml} files only, 14 more contained Kotlin code, one used web-based technologies and hence had no Java/Kotlin/resource files, and two bug reports with tangled commits (i.e., a large number of file changes) in the fixed commits.
We excluded these 33 bug reports resulting in 90 bug reports having at least one buggy Java file.  

During the collection process for the GUI Interaction data for the remaining 90 bug reports, were not able to collect GUI interaction data for 30 bug reports due to the one of following reasons: 
(1) the buggy behavior could not be reproduced; 
(2) bugs could not be reproduced due to the constraints of the step-by-step replay process (\eg logging into some apps was not possible); 
and (3) the step-by-step replay process has limitations including the inability to rotate the screen or execute fast swipe gestures. 
We excluded these 30 reports from our dataset, leaving 60 bug reports.

Finally, the two authors then worked separately to identify relevant buggy files reaching consensus in 53/60 ($\approx$88.33\%) of the cases (\ie the set of buggy code files matched). 
When there was no consensus, the three authors mutually finalized the buggy files.
In total, we identified the buggy files and GUI metadata containing screenshots, XMLs and event execution information for 80 bug reports (20 in the initial round and 60 in the subsequent round). {The collection of the GUI interaction data took $\approx$60-70 hours for the 80 bug reports originating from 39 apps (\ie $\approx$1.5 hours per app). However, it should be noted that in practice, the GUI interaction data that we collect in this study could be collected automatically through any number automated input generation tools for Android~\cite{kong2018automated,Fazzini2018,Zhao2019,zhao2022recdroid+,zhang2023automatically}.}

\subsection{Baseline Techniques}
\label{sec:baselineTechniques}

\subsubsection{\textbf{BugLocator}}
Zhou~\etal introduced \BugLocator~\cite{Zhou2012}, which uses a revised vector space model (rVSM) to obtain a ranked list of buggy files when a bug report is used as a query. 
Initially, a classic VSM based approach calculates the cosine similarity between the vector representations of the query and a document. 
\BugLocator then ranks longer documents higher, assuming that these files are more likely to contain bugs. 
\BugLocator is also capable of learning from previously fixed bugs by constructing a three-layer heterogeneous graph and computing a similarity score between past confirmed buggy files and the corpus of files under analysis. However, in this paper, we do not make use of this feature of \BugLocator, as we do not have the necessary data. 

\subsubsection{\textbf{Neural Embeddings via SentenceBERT}}
In the second baseline technique, we use the \SentenceBERT~\cite{reimers2019sentence} neural language model, which is a modification of a pre-trained \BERT model~\cite{Devlin2019}. 
\SentenceBERT augments the traditional \BERT model with siamese and triplet networks allowing for better support of tasks such as clustering and semantic search with less computational overhead. The model was fine-tuned on a popular natural language inference dataset~\cite{williams2018} and outperforms state-of-the-art approaches. 

We use the sentence transformer \textit{msmarco-distilbert-base-v3} implementation \cite{distilbert-huggingface} from the HuggingFace library \cite{huggingface} for our implementation of \SentenceBERT. 
The model uses 768-dimensional vector space and has a maximum sequence length of 510. 
As such, while all of our bug reports fit within this sequence length, certain source code files may exceed it. 
Thus, we split source code files into different segments with a maximum length of 510 tokens, and created an embedding for each segment. 
We compute cosine similarities between the bug report embedding and each segment of each source code file, and take the segment with the highest similarity value to the query as the similarity value for a given file, which we use for ranking the files.

\subsubsection{\textbf{Neural Embeddings via UniXCoder}}
In addition to neural language models trained on general natural language understanding tasks, we also wanted to explore how GUI information may complement document embeddings generated from a model trained primarily for \textit{code understanding} tasks. 
As such, for the third baseline, we use the \UniXCoder~\cite{guo2022unixcoder} model that is based on a multi-layer Transformer~\cite{Vaswani2017} architecture. We use the open-source implementation of \UniXCoder \cite{unixcoderImpl} to create embeddings of the source code and bug reports. 
The \textit{unixcoder-base} model is pre-trained on the CodeSearchNet \cite{Husain2019} dataset, one of the largest model training datasets for code understanding tasks containing two million code-comment pairs (across six programming languages). Similar to \SentenceBERT, this model also inputs a maximum token length, but of 512 opposed to 510. 
Therefore, we follow the same segmentation and similarity score calculation method as we do for \SentenceBERT, again using a cosine similarity measure.

\subsubsection{\textbf{Lucene}}
For the fourth baseline, we use \Lucene \cite{apacheLucene}, an open-source Java project that provides features to retrieve relevant documents. 
\Lucene uses a vector space model and \TFIDF document vector representations to rank buggy files based on an input query.

\subsubsection{\textbf{Preprocessing of Queries and Source Code Files}}
\label{sec:preprocessing}

We perform the following preprocessing steps, commonly used in past work on TR-based bug localization, for both queries (bug reports/reformulated queries) and source code files: splitting camel case and removing numbers, punctuation, tokens of length $1-2$, any special characters not part of English alphabets, and Java keywords. 
These steps were used for all techniques except source code for \BugLocator as this technique applies its own preprocessing~\cite{Zhou2012}. 
\looseness=-1

\subsection{Approach Configurations}
\label{subsec:configs}

\begin{table}[t]
\footnotesize
\caption{\small Combinations of the Filtering+Boosting re-ranking methods and GUI-Related File types considered. GS = GUI Screen; EGC = Exercised GUI Component; SC = GUI Screen Components.}
\label{tab:configs}
\begin{tabular}{P{1cm}|P{1cm}|P{1cm}|P{1cm}|P{1cm}|P{1cm}}
{\scriptsize \backslashbox{\textbf{Flt}}{\textbf{Bst}}}          & \textbf{GS} & \textbf{EGC} & \textbf{GS+EGC} & \textbf{SC} & \textbf{GS+SC} \\ \hline
\textbf{GS}                   &            &              &                              &           &                             \\ \hline
\textbf{EGC}                 &            &              &                              &           &                             \\ \hline
\textbf{GS+EGC} 				& \checkmark          &   {\checkmark}           &                              &           &                             \\ \hline
\textbf{SC}                    & \checkmark          & \checkmark            & \checkmark                            &           &                             \\ \hline
\textbf{GS+SC}  				& \checkmark          & \checkmark            & \checkmark                            & \checkmark         &                             \\ \hline
\end{tabular}
\vspace{-1.5em}
\end{table}

Our study has four main configuration parameters: 
(i) the number of GUI screens preceding the buggy screen to be used for GUI-Related File derivation (we investigate between 2-4 screens, including the buggy screen); 
(ii) the type of GUI interaction information used (we investigate the five combinations shown in Table~\ref{tab:gui-info});
(iii) the query reformulation techniques used (query replacement or expansion); 
and (iv) the re-ranking techniques used (filtering only, boosting only, and filtering + boosting). 
 
We examine \textit{\textbf{all}} combinations of these four parameters. 
However, one of our re-ranking techniques (filtering + boosting) is tightly coupled to the type of GUI information used, and as such we only investigate the feasible combinations {(\ie where boosted files are a subset of filtered files)} of the information shown in Table~\ref{tab:configs}. 
The reason we cannot explore all types of information is due to the fact that we cannot filter using a more restrictive set of GUI-Related Files (\eg GUI Screen-related Files) and boosting with a less restrictive set (\eg GUI Screen Component-related Files) as many (if not all) of the files filtered out would be the same that would then be subsequently boosted. 
In total, \textbf{we explore 657 configurations} of GUI Information for each baseline, for a \textbf{total of 2,628 configurations} across all of our experiments. {Table~\ref{tab:number-of-configs} shows the total number of configurations, where the number of configurations for each augmentation technique is calculated by multiplying feasible combinations of GUI types with the number of GUI screens.}

\begin{table}[t]
	\small
	\centering
	\caption{{Number of configurations. GIT = GUI Information Type; RRT = Re-ranking Technigues; QE = Query Expansion; QR = Query Replacement}}
	\label{tab:number-of-configs}
	\resizebox{0.98\columnwidth}{!}{%
	\begin{tabular}{P{1.4cm}|P{1.2cm}|P{1.0cm}|P{1.0cm}|P{0.8cm}|P{1cm}}
	\hline
		\textbf{ Augmentation} & \textbf{\# GIT~(RRT)}           & \textbf{\# GIT~(QE)} & \textbf{\# GIT~(QR)} & \textbf{\# Screens} & \textbf{$\times$ configs}\\
		\hline
		\multirow{3}{*}{\textbf{Filtering}} & 5 & & & 3 & 15\\ 	
		& 5 & 5 & &3 & 75\\
		 & 5 &  & 5 &3 & 75\\
		\hline	
		\multirow{3}{*}{\textbf{Boosting}} & 5 & & & 3 & 15\\ 	
		& 5 & 5 & &3 & 75\\
		 & 5 &  & 5 &3 & 75\\
		\hline	
		\textbf{Filtering} & 9 & & & 3 & 27\\ 	
		+ & 9 & 5 & &3 & 135\\
		 \textbf{Boosting} & 9 &  & 5 &3 & 135\\
		\hline	
		\textbf{Query}	& & 5 & & 3 & 15\\
		 \textbf{Reform.}& &  & 5 &3 & 15\\
		\hline	
             \multicolumn{5}{c|}{\textbf{Total Number of Configurations}} & \textbf{657}\\
		\hline

	\end{tabular}
	}
	\vspace{-1em}
\end{table}

\subsection{Metrics and Comparative Evaluation}
\label{sec:metrics}
We adopt Hits@K, a metric widely used in the literature~\cite{Zhou2012,rahman2017improved,chaparro2019using}, to evaluate the performance of augmenting our baseline techniques with GUI-related information.

\noindent{\textbf{Hits@K:}} This metric computes the percentage of queries for which a bug localizer retrieves at least one buggy file within the top-K files returned.
We report results for K=1,5,10 as past work has illustrated that the likelihood that a developer would look beyond 10 results is low~\cite{Wang2015}. We adopt this metric as it supports a practical scenario for bug localization. {Hits@K values fall in [0, 1], where higher values mean higher bug localization effectiveness.}

\noindent\textbf{Relative Improvement to Hits@10:} In addition to the Hits@K metric, given that the aim of our study is to compare the baseline techniques with their augmented counterparts, we also defined a comparative metric that measures the improvement of Hits@10 of one of our studied GUI Information configurations to a given baseline technique. 
This is defined as:

$$\frac{Hits@10GUI-Hits@10Base}{Hits@10Base}$$

\section{Empirical Results}
\label{sec:results}

In this section, we present the results of our empirical analysis organized by RQ. 
Of the 2,628 configurations of baseline techniques augmented with GUI interaction information, 1,080 configurations resulted in improvement over the baseline in terms of Hits@10, and 1,548 configurations resulted in no improvement or a degradation in effectiveness over the baselines. 
However, encouragingly, we find that a \textit{small set} of similar configurations of GUI-based augmentation methods tend perform best \textit{across all baseline techniques}, and more encouraging still, these best performing configurations result in marked improvements to Hits@10 (\eg up to 18\%) with little degradation to the ranks of buggy files already ranked within the top 10 by the respective baseline techniques. 
That is, for a small set of configurations that perform well across baselines, augmenting TR-based bug localization techniques with GUI interaction information provides a largely \textit{complementary} improvement in effectiveness.

\begin{table}[t]
	\setlength\tabcolsep{3.5pt}
	\small
	\centering
	\caption{{Number of configurations exhibiting positive \% improvement in Hits@10 over baselines across \# of screens.}}
	\label{tab:positive-improvement-configs}
	\begin{tabular}{c|c|c|c|c}
	\hline
		\textbf{Approach}           & \textbf{2 Screens} & \textbf{3 Screens} & \textbf{4 Screens} & \textbf{\# Screens}\\
		\hline
		BugLocator & 90 & 115 & 110 & 315\\
		SentenceBERT &	61 & 69 & 75 & 205\\
		UnixCoder &	101 & 114 & 97 & 312\\	
		Lucene & 80 & 86 & 82 & 248\\	\hline			
		\textbf{Total} & \textbf{332} & \textbf{384} & \textbf{364} & \textbf{1,080}\\	
	\hline
	\end{tabular}
	
	\vspace{-1em}
\end{table}

\begin{table}[t]
	\setlength\tabcolsep{3.5pt}
	\small
	\centering
	\caption{Average positive \% improvement of Hits@10 over baselines across the number of screens.}
	\label{tab:avg-positive-improvement}
	\resizebox{0.98\columnwidth}{!}{
	\begin{tabular}{c|c|c|c|c|c|c|c|c|c}
	\hline
		\textbf{Approach}           & \multicolumn{3}{c|}{\textbf{2 Screens}} & \multicolumn{3}{c|}{\textbf{3 Screens}} & \multicolumn{3}{c}{\textbf{4 Screens}}\\
		\hline
		 & \textbf{min} & \textbf{avg} & \textbf{max} & \textbf{min} & \textbf{avg} & \textbf{max} &
		 \textbf{min} & \textbf{avg} & \textbf{max}\\	\hline
		BugLocator & 1.75 & \textbf{7.86} & 17.54 & 1.75 & 7.78 & 17.54 & 1.75 & 7.74 & 17.54\\
		SentenceBERT &	1.72 & 5.51 & 12.07 & 	1.72 & 8.05 & 15.52 &	1.72 & \textbf{8.11} & 15.52\\
		UnixCoder &	1.79 & 7.28 & 14.29 & 	1.79 & \textbf{7.69} & 12.50 & 	1.79 & 6.76 & 14.29\\	
		Lucene & 1.56 & 5.55 & 9.37 & 1.56 & 5.65 & 10.94 & 1.56 & \textbf{5.79} & 12.50\\	\hline			
		Overall & 1.56 & 6.55 & 9.37 & 1.56& \textbf{7.29} & 10.94 & 1.56& 7.10 & 12.50\\	
	\hline
	\end{tabular}
	}
	\vspace{-1em}
\end{table}

\subsection{\textbf{RQ$_1$}: Impact of Number of Screens}
{The number of configurations across different numbers of screens that exhibit \textit{positive} \% improvement over baselines are shown in Table~\ref{tab:positive-improvement-configs}.}
Table~\ref{tab:avg-positive-improvement} reports the average, minimum, and maximum \textit{positive} improvement over each respective baseline technique across \textit{all} studied GUI Information configurations when different numbers of screens from the bug reproduction scenario are used.   
Given that we want to understand which screen configuration provides the best \textit{improvement}, here we do not discuss configurations that did not improve over the baseline. 
The highest \textit{average} improvement for each baseline technique is shown in bold. 
From the table we can observe that there is no one configuration for number of screens that performs best across all techniques. 
However, considering average overall improvement, using information from 3 screens (\ie the buggy screen and two prior screens) provides the highest overall improvement over the baseline techniques, whereas using information from 4 screens provides the largest improvement for two techniques (Lucene and SentenceBERT). {It should be noted that 4 screens leads better performance in terms of the \textit{max improvement} of a single configuration over the baselines. However, the average values correspond to the highest increase across \textit{all} studied configurations.} 

\begin{tcolorbox}
\small
\vspace{-0.5em}
\textbf{Summary of Findings for RQ$_1$:} We find that using GUI information from the buggy screen, and two preceding screens provides the highest overall increase in effectiveness across our studied baseline TR-based bug localization techniques. 
This indicates that relevant GUI information for bug localization is contained in not only in the buggy screen, but also in the preceding screens.
\vspace{-0.5em}
\end{tcolorbox}

\subsection{\textbf{RQ$_2$}: Impact of GUI Information Type \& Augmentation Method}

\begin{table}[t]
	\setlength\tabcolsep{3.5pt}
	\small
	\centering
	\caption{Best performing GUI Information configurations according to \% relative improvement for Hits@10.}
	\label{tab:best-config-diff-augmentation}
	\resizebox{0.98\columnwidth}{!}{%
	\begin{tabular}{c|c|c|c|c}
	\hline
		\textbf{Approach}           & \textbf{Augmentation} & \textbf{Information Type} & \textbf{\# Screens} & \textbf{HIT@10 Improvement} \\
		\hline
					
		& Filtering	 & SC & 3 & 1.75 \\
		 & Boosting &	GS & 4 & 12.28\\
		BugLocator & Filtering+Boosting &	SC(F)+GS(B) & 4 & 14.04\\	
		& Query Expansion &	SC & 2 & 8.77\\ 
	    & Query Replacement &	GS+SC  & 2 & 1.75\\ \hline
		 
		& Filtering	 & SC & 2 &  5.17 \\
		 & Boosting &	GS+EGC & 3 & 15.52\\
		SentenceBERT & Filtering+Boosting &	SC(F)+[GS+EGC(B)] & 3 & 15.52\\	
		& Query Expansion &	--- & --- & ---\\ 
	    & Query Replacement & ---	& --- & --- \\ \hline
		 
		 & Filtering & --- & --- & --- \\
		 & Boosting &	GS & 4 & 12.50\\
		UniXCoder & Filtering+Boosting &	SC(F)+GS(B) & 4 & 12.50\\
		& Query Expansion &	SC & 2 & 14.29\\ 
	    & Query Replacement &	GS+SC & 3 & 8.93\\	\hline
		 
		 & Filtering	 & SC & 3 & 4.69 \\
		 & Boosting &	GS & 4 & 9.38\\
		Lucene & Filtering+Boosting &	SC(F)+GS(B) & 4 & 12.50\\
		& Query Expansion &	GS & 4 & 6.25\\ 
	    & Query Replacement &	--- & --- & ---\\
	\hline
	\end{tabular}
	}
	\vspace{-1em}
\end{table}

In RQ$_2$, we first aim to investigate the impact of using GUI-related files sourced from different types and combinations of GUI-Interaction Information (\eg GUI Screens (\texttt{\small \textbf{GS}}), Exercised GUI Components (\texttt{\small \textbf{ECG}}), \texttt{\small \textbf{GS+ECG}}, Screen Components (\texttt{\small \textbf{SC}}), and \texttt{\small \textbf{GS+SC}}) on our augmentation methods.  
To do this, in Table~\ref{tab:best-config-diff-augmentation}, which is organized by augmentation method, we report the best performing GUI Interaction Information type for each of our five augmentation methods and for each baseline technique. 
This allows us to examine whether there are trends in the best performing information types across our studied techniques. 
Dashes signify that no configuration of the reported augmentation method improved over the baseline. 
Table~\ref{tab:best-config-diff-augmentation} illustrates a few notable trends across the different augmentation techniques. 
For instance, we can observe that for \textit{Filtering}, \texttt{\small SC} (which map to the largest number of GUI-related files) is always the best performing information type (except for UniXCoder). 
This is not entirely surprising as, due to the large number of associated GUI-related files, it is the least restrictive filtering. 

For \textit{Boosting}, \texttt{\small GS} (or \texttt{\small GS+ECG}) performs best across all baseline techniques. 
Given that \texttt{\small GS} information targets a smaller number of GUI-related files that often encompass large portions of screen functionality, it follows that boosting such targeted information is likely to have a larger positive effect on ranking buggy files. 
These same trends also hold for \textit{Filtering+Boosting}, wherein filtering with GUI Interaction Information that is mapped to a larger number of GUI-related files (\eg \texttt{\small SC}) and boosting with a more targeted information type that maps to fewer files (\eg \texttt{\small GS}, \texttt{\small ECG}) generally leads to the best results. Finally, we find that \textit{Query Replacement} generally does not lead to an improvement over the baseline, but Query Expansion does lead to improvements -- however, there is no clear trend in the best performing Information type for expansion.

\begin{figure}[t]
		\vspace{-1em}
		\centering
		\includegraphics[width=\linewidth]{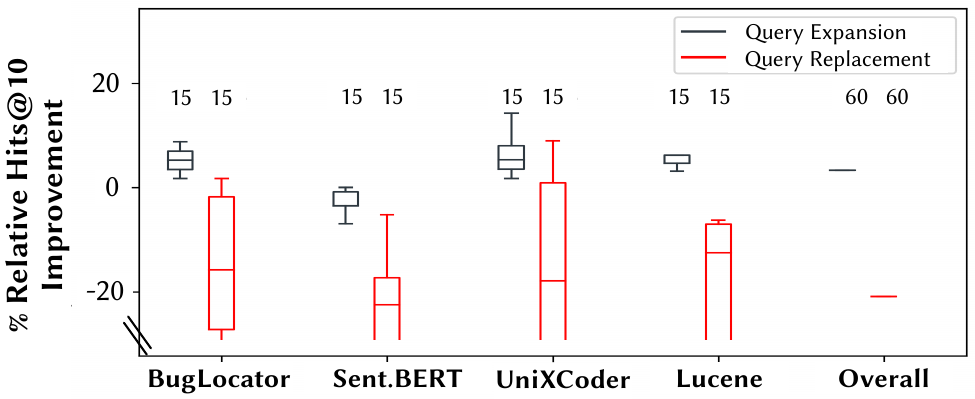}
		\vspace{-1em}
		\caption{Relative \% Improvement of Query Reformulation}
		\label{fig:gui-query-reformulation-relative-improvement}
		\vspace{-1em}
\end{figure}

Turning our attention to the impact of different augmentation methods we observe further confirmation of trends that began to surface when examining the impact of different GUI-Interaction Information types. 
First, we examine the reformulation and re-ranking methods in isolation to better understand their effects. Figure~\ref{fig:gui-query-reformulation-relative-improvement} shows a box-and-whisker plot illustrating the relative improvement that both query expansion and replacement have over our studied baseline techniques. 
The number of configurations for each augmentation method are given above each plot. 
This plot shows that, in general, Query Replacement rarely results in any improvement over the baseline technique. 
On the contrary, query expansion does generally result in improvement for every baseline except SentenceBERT.
This is likely due to the abstract semantic embeddings that are produced by SentenceBERT which may be able to more naturally retrieve GUI-related information due to its learned, rich term similarities. Furthermore, query expansion seems to have the largest positive effect on UniXcoder, which is another neural model trained on code as opposed to natural language. We speculate that this improvement may be due to the fact that embedding file names into UniXCoder's code-specific embedding space makes it far more likely for those files to be retrieved as compared to SentenceBERT. The observation that query replacement tends to perform poorly indicates that there is often important lexical information contained within bug reports for TR-based localization techniques, as removing this information and replacing it with GUI-related file names generally degrades performance (sometimes markedly so). Conversely, \textit{expanding} the bug report query with GUI-related terms does appreciably improve most techniques, signaling that expanding queries with GUI-related file names is helpful.

\begin{figure}[t]

		\centering
		\includegraphics[width=\linewidth]{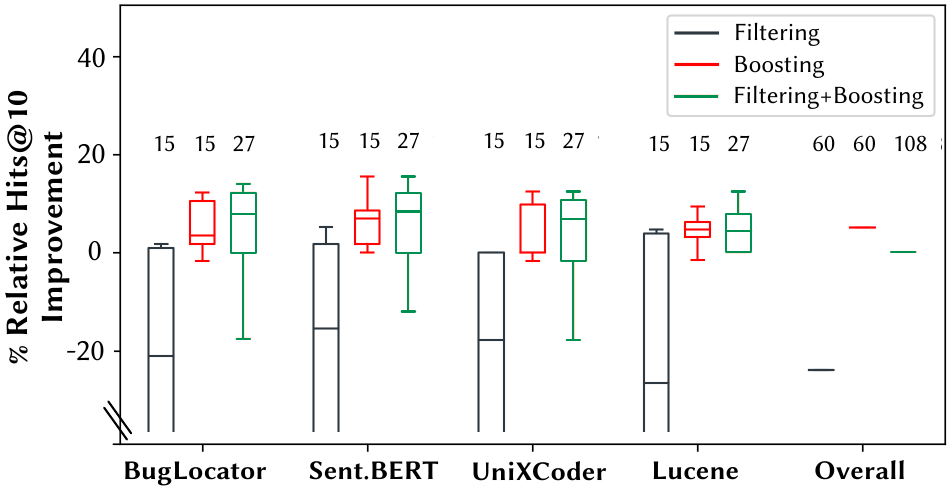}
		\vspace{-1em}

		\caption{Relative \% Improvement of  Re-Ranking}
		\label{fig:gui-augmentation-relative-improvement}
		\vspace{-1em}
\end{figure}

Figure~\ref{fig:gui-augmentation-relative-improvement} illustrates the same box and whisker plot, but for our three re-ranking methods. Form this plot we can observe that only a small number of filtering configurations improve upon the baseline techniques. This follows from our findings in RQ$_2$ which illustrate that only filtering configurations with a GUI information type that maps to a large number of GUI-related files perform well, as this limits the overall number of files filtered out of the searchable corpus. However, while filtering only sometimes leads to improvements, nearly every configuration of Boosting leads to appreciable improvements over the baseline techniques. Combining Filtering and Boosting together leads to the highest overall median improvements, but with more variability in the results due to poor performance of filtering with certain information types. The results related to Boosting illustrate that re-ranking GUI-related files to the top of list of retrieved results is nearly always beneficial, which indicates that these files have a far higher probability of being buggy compared to other files that were not linked to GUI information.

\begin{tcolorbox}
\small
\vspace{-0.5em}
\textbf{Summary of Findings for RQ$_2$:} We find that for Filtering, \texttt{\footnotesize SC} information leads to the largest improvement, and for Boosting, \texttt{\footnotesize GS} information leads to the largest improvement. 
Overall, filtering with information types which map to a higher number of GUI-related files (\ie filtering out fewer files), and boosting with information types that map to a lower number of files, tend to perform best. 
Filtering only provides benefit in a small number of cases with limited GUI Information types, whereas boosting is nearly always beneficial.
We also observe that Query Replacement rarely leads to performance improvements, whereas Query Expansion does lead to improvement for every baseline except SentenceBERT, albeit without any single GUI Information type performing best. Combining Filtering and Boosting together leads to the highest overall improvements, but these configurations are more sensitive to the information types used.
\vspace{-0.5em}
\end{tcolorbox}

\begin{table}[t]
	\setlength\tabcolsep{3.5pt}
	\small
	\vspace{-0.5em}
	\centering
	\caption{ Best Performing Combinations}
	\label{tab:best-performing-combinations}
\resizebox{0.98\columnwidth}{!}{%
	\begin{tabular}{c|c|c|c|c|c|c|c}
	\hline
		\textbf{Baseline/}  & \textbf{Filtering} & \textbf{Boosting} & \textbf{GUI Info} & \textbf{\#} & \textbf{H@1} & \textbf{H@5} & \textbf{\#Bugs Top10}\\ 
		\textbf{Config} & \textbf{GUI Info} & \textbf{GUI Info} & \textbf{Query Exp.} & \textbf{Scrns} & & & \textbf{(H@10)}\\
		\hline

		\multirow{2}{*}{BugLocator}	& None & None & None & & 0.39 & 0.60 & 57 (0.71)\\
		 & SC & GS & GS+SC & 3 & 0.33 & 0.76 & 67 (0.84)\\ 
		\hline
	
		\multirow{5}{*}{SentenceBERT}	& None & None & None & & 0.23 & 0.56 & 58 (0.72)\\
		
		 & SC & GS+EGC & None & 3 & 0.30 & 0.72 & 67 (0.84)\\ 
		 
		 & GS+SC & GS+EGC  & None & 3 & 0.30 & 0.72 & 67 (0.84)\\
		
		 & SC & GS+EGC & EGC & 3 & 0.30 & 0.72 & 67 (0.84)\\
		
		 & GS+SC & GS+EGC & EGC & 3 & 0.30 & 0.72 & 67 (0.84)\\
		
		\hline
									
		\multirow{3}{*}{UnixCoder}	& None & None & None & 	& 0.14 & 0.62 & 56 (0.70)\\
		& SC & GS & SC & 4	& 0.31 & 0.75 & 64 (0.80)\\
		
		& GS+SC & GS & SC & 4 & 0.31 & 0.75 & 64 (0.80)\\ \hline					
									
		\multirow{3}{*}{Lucene} & None & None & None & & 0.40 & 0.75 & 64 (0.80)\\
		& SC & GS & None & 4 & 0.36 & 0.80 & 72 (0.90)\\
		
		& GS+SC & GS & None & 4 & 0.36 & 0.80 & 72 (0.90)\\
												
	\hline
	\end{tabular}
}

\end{table}

\begin{table*}[t]
	\setlength\tabcolsep{3.5pt}
	\small
	\centering
	\caption{In-Depth Analysis of Best Performing overall GUI-related augmentation configuration.}
	\label{tab:qual-analysis}
	\resizebox{0.98\textwidth}{!}{%
	\begin{tabular}{c|c|c|c|c|c|c|c|c|c|c|c}
	\hline
		\textbf{Baseline/}           & \textbf{Filtering} & \textbf{Boosting} & \textbf{Number of} & \textbf{Hits@10} & \textbf{\# Bugs Out10} & \textbf{\# Bugs In10} & \textbf{\# Bugs Inside } & \textbf{\# Bugs Inside} & \textbf{\# Bugs Inside} & \textbf{\# Bugs Outside } & \textbf{\# Bugs Outside }\\ 
		\textbf{Config} & \textbf{GUI Info} & \textbf{GUI Info}  & \textbf{ Screens} & & \textbf{ $\rightarrow$ In10} & \textbf{ $\rightarrow$ Out10}& \textbf{10 Improved} & \textbf{10 Deteriorated} & \textbf{10 Unchanged} 
		& \textbf{10 Improved} & \textbf{10 Deteriortated}\\
		\hline
		
		BugLocator & SC & GS & 4 & 0.81 & 8 &	0	& 16	& 21	& 20& 	10&	4\\
		SentenceBERT & SC & GS & 4 & 0.81 & 7	&0	&22	&15&	21	&12	&1\\
		UniXCoder & SC & GS & 4 & 0.79 & 7&	0&	21	&13	&22	&15	&0\\
		Lucene & SC & GS & 4 & 0.90 & 8	&0	&16	&24	&24	&6	&1\\ 
		\hline
	\end{tabular}
	}
	\vspace{-1em}
\end{table*}

\subsection{\textbf{RQ$_3$}: Best Performing Configurations}

RQ$_3$ aims to take a deeper look at the best performing individual GUI-augmentation configurations for each baseline technique. 
We report this information in Table~\ref{tab:best-performing-combinations} where the best performing configurations were selected by taking the top performing techniques according to Hits@10, and breaking ties according to Hits@5. 
If there were still ties after considering Hits@5, then we report all such configurations. 
The first major observation that can be made from Table~\ref{tab:best-performing-combinations} is that \BugLocator and \Lucene, which use more traditional text-retrieval baselines (\ie \TFIDF document representations) tend outperform the techniques that use neural embeddings (SentenceBERT and UniXCoder). 
It should be noted that, given the goal of our study is to examine the benefit of GUI interaction information, we used \SentenceBERT and \UniXCoder in a zero-shot setting, wherein they were not fine-tuned on bug report information. 
Future work may examine these in a fine-tuned setting. 

The next major trend illustrated in Table~\ref{tab:best-performing-combinations} is that the best performing configurations offer a marked improvement (in terms of Hits@5 and Hits@1-) compared to the baseline techniques, with relative improvements ranging from 12.5\%-18\% for Hits@10, and 6\%-29\% for Hits@5. 
This means that, for all baselines, the best performing configurations result in the inclusion of buggy files for an additional 9-10 bugs in the top 10 results. 
Furthermore, we find that the best performing configurations of our GUI-augmentation methods are strikingly consistent. 
That is, Filtering+Boosting is always the best re-ranking technique, using \texttt{\small SC} or \texttt{\small GS+SC} GUI information for Filtering, and either \texttt{\small GS} or \texttt{\small GS+EGC} information for boosting always leads to the largest improvements over the baseline techniques. 
This means that any one of these configurations can likely be applied to a TR-based baseline and improve the results.
We find that 
certain baselines and configurations seem benefit from query expansion with varying GUI information types. 
This is consistent with the observations from the previous RQ.
Finally, we performed a Wilcoxon signed rank test on the first rank of a buggy file for all bugs (even those ranked outside top-10) across all techniques using a 95\% confidence interval. 
We found that all of the best performing GUI-augmentation methods outperform their corresponding baseline techniques to a statistically significant degree, save for \BugLocator, which had a p-value of 0.44. 
However, given that past work has illustrated that TR-based bug localization typically only provides benefits to developers if the buggy files is ranked within the top-10 results~\cite{Wang2015}, we must note that \BugLocator saw the benefit of the largest overall increase in Hits@10 from our GUI-augmentation methods across all of our studied baselines.

\begin{tcolorbox}
\small
\vspace{-0.5em}
\textbf{Summary of Findings for RQ$_3$:} We find that the best performing configurations of our GUI-augmentation methods combine filtering and boosting, using \texttt{\footnotesize SC} or \texttt{\footnotesize SC+GS} to filter, and \texttt{\footnotesize GS} or \texttt{\footnotesize GS+EGC} to boost, and lead to an improvement in Hits@10 ranging from 12.5\%-18\%, and 6\%-29\% for Hits@5.
\vspace{-0.5em}
\end{tcolorbox}

\vspace{-1em}
\subsection{Discussion}

We aim to see whether a single GUI augmentation configuration can perform well across all baselines. 
To this end, we first ranked our 657 configurations for each baseline in terms of their performance in Hits@10, using the average performance of HITS@1 and HITS@5 to break ties. 
We found two configurations that perform identically, and chose the simpler configuration of the two to analyze. 
This configuration uses the Filtering+Boosting 
with \texttt{\small SC} information for Filtering and \texttt{\small GS} information for boosting, using data from 4 screens.
\looseness=-1

In general, an augmentation affects the rankings of most of the bugs, however we are less concerned with those that are far from top-10 (they are still hard to retrieve), or those already in top-10 (they are still easy to retrieve).
We consider an augmentation to be good if it ranks more bugs in top-10 than its baseline (\ie at least one of their buggy files rank in top-10), in other words bugs that were hard to retrieve with the baseline are now easy to retrieve with the augmentation.
With that in mind, for the GUI augmentation method configuration we identify as best, we examined how it performed in terms of moving bugs from outside the top 10 ranks to inside the top 10 ranks (and vice versa) as well as the number of buggy file that improved, degraded and remained unchanged both inside and outside the top 10 ranks -- we report these results in Table~\ref{tab:qual-analysis}. 
This table illustrates that the best performing GUI-augmentation configuration brings buggy files for 7-8 bugs from outside the top-10 to inside the top-10 without moving any buggy files that were previously in the top 10 outside of the top 10. 
This signals that the GUI-related augmentations are largely complementary to existing techniques and do not hurt existing bugs ranked within the top-10. 
When examining how ranks of buggy files change within the top 10 ranks, we find a relatively even mix of files that improved (16-22), deteriorated (13-24), and remained unchanged (20-24). This signals that there is limited net impact on the ranks of the files within the top-10 ranks. 
Furthermore, we find that this configuration improves more buggy files outside the top-10 than it deteriorates, and our studied configuration outperform \textbf{all} baselines in terms of top ranks of buggy files according to a Wilcoxon signed rank test at a 95\% confidence interval.

\begin{figure}[t]
		\centering
		\includegraphics[width=\linewidth]{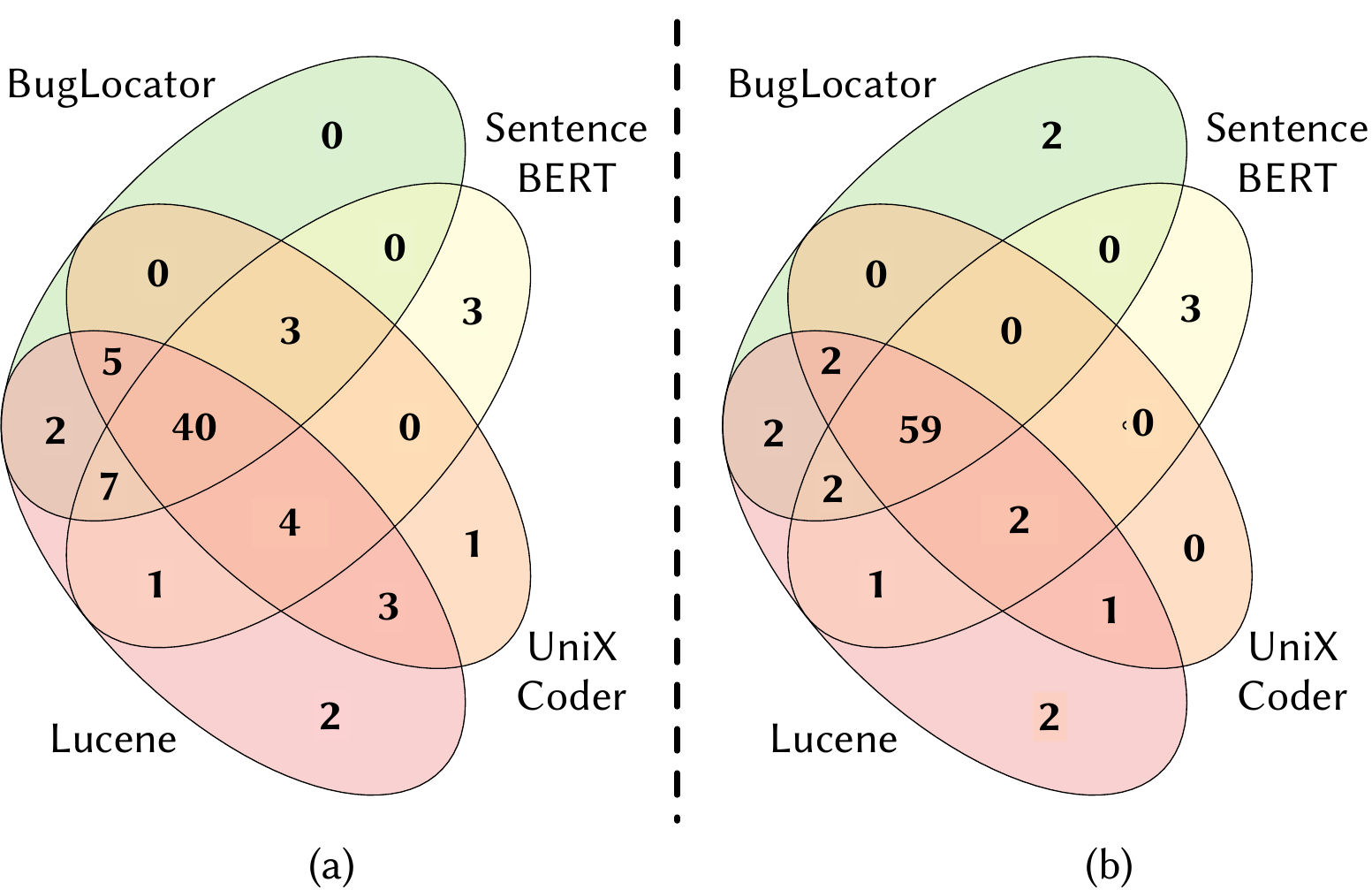}
		\caption{{Venn Diagram illustrating the overlap where buggy files appear in the top-10 results for our studied baselines (a) and for the best performing GUI augmentation methods (b).}}
		\label{fig:venn-diagram}
		\vspace{-2em}
\end{figure}

{In addition, we wanted to understand the degree of orthogonality, in terms of Hits@10, across our baseline techniques, and how our GUI augmentation methods effect this orthogonality. In Figure~\ref{fig:venn-diagram}-a, we provide a Venn Diagram that includes each baseline, where overlapping sections indicate a shared bug for which a buggy file was ranked within the top-10. From this diagram, we can observe that between 61\% and 70\% of bugs with buggy files ranked within the top 10 are shared across all baselines (40 bugs), indicating that the different approaches do exhibit some degree of orthogonality, with Lucene having the highest amount of orthogonality in comparison to the other techniques. However, taking the best performing GUI-augmented configuration for each baseline decreases the overall amount of orthogonality, as between 92\% and 82\% of bugs with buggy files ranked within the top-10 are shared. \textit{This indicates that even when we take baseline techniques that operate using different text representations and exhibit orthogonality, our GUI-augmentation methods are able to improve upon all of them despite their differences.}}

\section{Threats to Validity}
\label{sec:threats}

\noindent{\textbf{Construct Validity:}} 	Our study's main threat to construct validity is rooted in the potential subjectivity of deriving our manually labeled bug localization dataset. To minimize this threat, we performed a rigorous manual labeling procedure (see Sec.~\ref{sec:coding}), wherein two authors independently coded the buggy files for each bug report, and conflicts were resolved via discussion with a third participant. All authors involved in this process were experienced in Android development and achieved a high inter-coder agreement (88.33\%).

\noindent{\textbf{Internal Validity:}} Threats to internal validity of our study conclusions stem from our selected baseline TR-based bug localization techniques. We observed a variation in effectiveness across our four studied TR-based localization baselines, however, despite this differing performance, the main commonality in our experiments was the introduction of our GUI-based augmentation techniques, which were the main factor in the improvements we observed. 

\noindent{\textbf{Conclusion Validity:}} To mitigate threats to conclusions, we made use of the common Hits@K metric, focusing our analysis on Hits@10. This choice was motivated by prior work that illustrated that automated bug localization techniques are generally only helpful to developers when they return relevant buggy files in the top 10 results~\cite{Wang2015}. Furthermore, in addition to examining effectiveness in terms of Hits@10, we also performed a fine-grained analysis into the impact that the GUI-based augmentation techniques have on the number of buggy files that get moved into/out of the top 10. 

\noindent{\textbf{External Validity:}} To mitigate threats to external validity we studied 2,628 combinations of our GUI-based augmentation methods, for 80 bug reports from 39 diverse, and popular applications. Given the effort required to build our dataset, the amount of data we analyze is sizable, but a larger set of reports would provide additional confidence in our results. We analyzed four TR-based bug localization techniques, and our results may not generalize beyond these.

\vspace{-1em}
\section{Related Work}
\label{sec:related_work}

\textbf{Text-Retrieval-based Bug Localization (TRBL) Techniques.}  A TRBL retrieval engine, powered by a TR technique, typically leverages the textual similarity between the code and the bug report to the determine the relevance of a code artifact to a query (less/more likely to be buggy). The relevance score is  determined using different techniques and representations of artifacts and reports.
\looseness=-1

Researchers have proposed a variety of TRBL techniques over the past two decades \cite{akbar2020large}, spanning Information Retrieval (IR) techniques \cite{kuhn2007semantic, effectiveness_of_IR,gay2009use,poshyvanyk2007feature,Rao2011,Rao2011,marcus2004information,Koyuncu2019DCAD} and, more recently, Deep Learning (DL)-based approaches \cite{xiao2017improving, wang2020multi, Huo2019, xiao2019improving,fang2021classification, yang2021locating,ciborowska2022fast}. Common IR techniques include classical algorithms such the Unigram Model (UM) \cite{Rao2011}, Cluster Based Document Model (CBDM) \cite{Rao2011}, Vector Space Model (VSM) \cite{Zhou2012, Yang2021}, Latent Semantic Indexing (LSI) \cite{marcus2004information}, and Latent Dirichlet Allocation
(LDA) \cite{lukins2008source}. DL-based approaches include models such as CNNs \cite{xiao2017improving, wang2020multi, Huo2019, xiao2019improving}, RNNs \cite{fang2021classification, yang2021locating}, Transformers \cite{ciborowska2022fast} and combinations of these \cite{xiao2018improving, yang2021utilizing, xiao2018bug}. Other approaches combine IR and DL techniques \cite{Lam2017}. The advantage of IR techniques is their simplicity and efficiency, since they do not require training. Conversely, DL-based approaches are more expensive as they require training using a large amount of labeled data. The advantage of DL techniques is they can generate more informative document representations. To assess the effect of a technique on TRBL when leveraging GUI interaction data, we used two IR-based techniques, namely BugLocator \cite{davies2012using, davies2013bug} and Lucene \cite{apacheLucene}, and two DL models, namely SentenceBERT \cite{reimers2019sentence} and UniXCoder~\cite{guo2022unixcoder}.

\textbf{Information Sources for TRBL.} To address the lexical gap between bug reports and source code and produce more accurate suggestions of buggy artifacts~\cite{Bettenburg2008GoodBR, ye2014learning, mills2020relationship}, researchers have utilized additional information related to the bug reports and code. The most prominent sources of information used by existing approaches include similar bug reports \cite{wong2014boosting, rath2018analyzing}, code structure \cite{wang2016amalgam+, youm2017improved, takahashi2018preliminary}, code version history \cite{Wang2014, wang2016amalgam+, youm2017improved, zhang2019commit},  stack traces \cite{Moreno2014, youm2017improved, Wen2016}, part-of-speech information \cite{zhou2017augmenting}, and combinations of the above \cite{wong2014boosting, shi2018comparing}. 

Particularly relevant is the work that leverages system execution traces for TRBL \cite{Moreno2014}, however, collecting these traces are typically expensive since they require program instrumentation or a suite of test cases with high-coverage~\cite{Le2015}.  We study how GUI interaction data of Android apps can be used to help bridge the lexical gap between bug reports and code, and thus help retrieve relevant buggy files more effectively. To the best or our knowledge, we are the first to explore the use of such data for TRBL. This data can be easily collected via the Android infrastructure without requiring app instrumentation \cite{androidSystemTracing}, when the apps are used or the bugs are reproduced, manually or via automated execution tools \cite{Moran2016}. 
\looseness=-1

\textbf{Re-ranking Methods for TRBL.} Prior work has used the aforementioned sources of information to produce a ranking of code artifacts or to re-rank them based on an existing ranking. The goal is to rank the buggy artifacts higher than non-buggy artifacts. We identify at least three general re-ranking methods used in the literature: \textit{boosting}, \textit{filtering}, and \textit{query (re)formulation}. 
Prior techniques have \textit{boosted} the similarity/relevance score of code artifacts to improve their ranking \cite{lou2021boosting, wong2014boosting,rocchio1971relevance}. 
Other techniques \textit{filter} out irrelevant code artifacts from the artifact search space or initial ranking \cite{liu2007feature}. 
Query (re)formulation aims to encode relevant textual information in the query for improving retrieval~\cite{florez2021combining,chaparro2019using, chaparro2017using, mills2020relationship, rahman2021forgotten,chaparro2019reformulating,chaparro2016reduction}. Researchers focused on three main reformulation methods: query expansion~\cite{carpineto2012survey, kim2021novel} (which adds extra terms to the query), query replacement~\cite{gibiec2010towards, guo2017tackling} (which substitutes the query with a new one), and query reduction~\cite{rahman2017improved, Rahman2018,chaparro2016reduction,florez2021combining} (which identifies/removes terms in the query that hinder retrieval). Inspired by this work, we assess the effect of using GUI interaction information on TRBL performance via boosting, filtering, query expansion, and query replacement. Our future work will explore how query reduction along with GUI information can be combined to improve TRBL performance.

\textbf{Software Types for TRBL.} The vast majority of prior work has proposed system-agnostic TRBL approaches \cite{Zhou2012, lukins2008source, poshyvanyk2009using}. This means they were designed to operate on any kind of software system (\eg desktop applications, libraries, and command line projects) and any bug report, without differentiating the bug type. Recent work has target deep learning projects~\cite{Kim2022}. Our study focuses on Android apps and GUI-based bugs, which represents the majority of bug types found in the Android ecosystem \cite{bhattacharya2013empirical, song2022toward,song2022burt}. 
\looseness=-1

\vspace{-0.5em}
\section{Conclusion \& Future Work}

We reported an empirical study that found a high positive effect of mobile app GUI interaction data on text-retrieval-based bug localization, using a manually-constructed bug localization dataset consisting of 80 bug reports with GUI metadata and four bug localization baseline techniques. The measured effect indicates GUI interaction data can help bridge the lexical gap between bug reports and source code, resulting in better rankings of buggy files suggested to developers. Our future work will investigate other ways of augmenting existing localization techniques via GUI interaction data for mobile apps and other types of systems and conducting human studies that aim to validate the benefits of using GUI interaction information for bug localization in practice.

%-------------------------------------

\section*{acknowledgements}

This work is supported by the U.S.\ NSF under grants: CCF-1910976, CCF-2343057, CCF-1955837, CCF-2007246, and CCF-1955853. Any opinions, findings, and conclusions expressed herein are the authors and do not necessarily reflect those of the sponsors.

\balance

\bibliographystyle{ACM-Reference-Format}
\bibliography{references}

\end{document}